\documentclass[11pt,amsmath,amssymb,
 aps,nofootinbib]{revtex4-1}
\usepackage{geometry}                
\usepackage{graphicx}
\usepackage{color}
\usepackage{xcolor}
\usepackage{stmaryrd}
\usepackage{datetime2}

\usepackage{amssymb}
\usepackage{amsthm}
\usepackage{amsfonts}
\usepackage{amsmath}
\usepackage{enumerate}
\usepackage{hyperref}

\usepackage{tikz}
\usepackage{simpler-wick}

\DeclareMathOperator{\Tr}{Tr}
\DeclareMathOperator{\tr}{tr}
\usepackage{epstopdf}
\theoremstyle{definition}

\newcommand{\ldb}{\llbracket}
\newcommand{\rdb}{\rrbracket}
\newcommand{\NC}{\mathrm{NC}}
\newcommand{\DP}{\mathrm{DP}}

\newcommand{\disavg}[1]{\mathbb{E}\left[ #1 \right]}

\DeclareMathOperator{\Wg}{Wg}

\begin{document}

\title{Many-body systems with random spatially local interactions}

\author{Siddhardh C. Morampudi}
\author{Chris R. Laumann}
\address{Department of Physics, Boston University, Boston, MA 02215, USA}

\begin{abstract}
    We extend random matrix theory to consider randomly interacting spin systems with spatial locality. 
    We develop several methods by which arbitrary correlators may be systematically evaluated in a limit where the local Hilbert space dimension $N$ is large. 
    First, the correlators are given by sums over \emph{stacked} planar diagrams which are completely determined by the spectra of the individual interactions and a dependency graph encoding the locality in the system.
    We then introduce \emph{heap freeness} as a generalization of free independence, leading to a second practical method to evaluate the correlators.
    Finally, we generalize the cumulant expansion to a sum over \emph{dependency partitions}, providing the third and most succinct of our methods.
    Our results provide tools to study dynamics and correlations within extended quantum many-body systems which conserve energy.
    We further apply the formalism to show that quantum satisfiability at large-$N$ is determined by the evaluation of the independence  polynomial on a wide class of graphs. 
\end{abstract}

\maketitle
\section{Introduction}

Consider a system of qudits ($N$-level degrees of freedom) interacting with each other through generic interactions. Is it possible to systematically calculate physical quantities in such a system? If the qudits are non-interacting, then all correlation functions can be trivially calculated. At the other extreme, if all qudits are acted upon by a non-local interaction, then many properties of the system are well described by random matrix theory. 

In this paper, we develop systematic methods to evaluate trace correlators in randomly interacting many-body systems with  spatial locality, when the dimension $N$ of the qudits is large. 
We consider a physical system consisting of $n$ interacting qudits%
\footnote{The Hilbert space has dimension $N^n$. The generalization to varying qudit dimensions $N_q$ is straightforward, so we use $N_q = N$ throughout to simplify the presentation.}
whose Hamiltonian is given by
\begin{equation}
\label{eq:generalham}
H = \sum_{i}^{M} O^i
\end{equation}
Here, the superscript $i$ runs over $M$ interactions, each of which acts on a subset $\partial i$ of the qudits%
\footnote{We use superscripts $i,j$ to label interactions and Greek subscripts $\alpha,\beta, \mu,\nu$ to label states in Hilbert space.}.
Each $O^i$ has a fixed spectrum represented by a diagonal matrix $\lambda^i$ which is rotated by Haar random unitaries $U^i$ acting  on the local Hilbert space defined by the $\partial i$ qudits, i.e, $O^i = U^{i}\lambda^i U^{i\dagger} \otimes \mathbb{I}_{n-k_i}$ where $k_i = |\partial i|$ is the degree of the interaction. 
The primary correlators of interest are the disorder averaged trace moments
\begin{align}
\label{eq:trace_moment_def}
    \ldb \cdots \rdb \equiv \frac{1}{N^n} \disavg{\Tr (\cdots) } \equiv \disavg{\tr (\cdots)}
\end{align}
where $\disavg{\cdots}$ denotes the average over the Haar unitaries $U^i$, and $\cdots$ denote general products of the operators $O^i$ \footnote{With suitable treatment of $e^{-\beta H}$ and/or $e^{-i H t}$, such trace moments include finite temperature and/or dynamical correlators.}.
To obtain a non-trivial large-$N$ limit, we take the eigenvalues of $O^i$ to be $O(N^0)$ so that the lowercase trace $\tr O^i$ is also $O(N^0)$. 

\begin{figure}[tbp]
\begin{center}
\includegraphics[width=\textwidth]{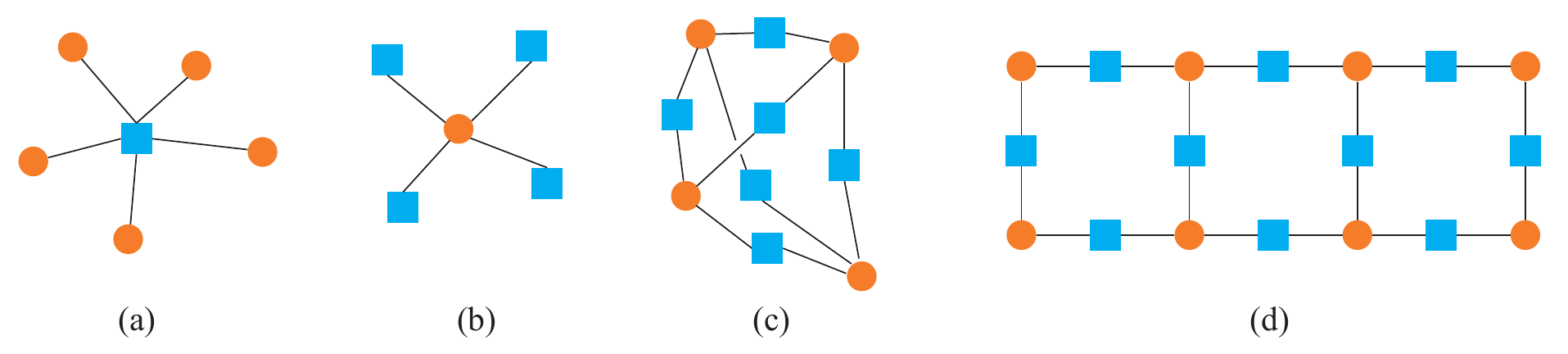}
\caption{
Different locality structures represented by interaction graphs $\mathcal{G}$ where circles represent qudits $q$ and squares the interactions $O^i$. 
(a) Five qudit degrees of freedom interacting through a single all-body interaction. This is the setting for classical random matrix theory.
(b) Four independent Hamiltonian operators acting on a single qudit -- the setting for free probability theory. 
(c) Four qudits interacting pairwise (2-local) and all-to-all.
(d) Eight qudits interacting pairwise (2-local) on a piece of square lattice, thus having spatial locality
}
\label{fig:intGraphs}
\end{center}
\end{figure}

The locality structure of Hamiltonians like $H$ can be represented by an interaction graph $\mathcal{G}$ specifying the qudits $q$ on which each interaction $O^i$ acts.
At one extreme, random matrix theory describes zero-dimensional many-particle systems such as nuclei and quantum dots by assuming that all of the degrees of freedom interact so strongly that the Hamiltonian can be considered as a single random matrix on the full Hilbert space\cite{Wigner1955, Mehta2004random, Guhr1998}. 
This corresponds to a star-like interaction graph as in  Fig.~\ref{fig:intGraphs}a. 
Free probability theory extends this zero-dimensional setting by allowing multiple independent random interaction terms to act on all of the degrees of freedom\cite{Voiculescu1991, NicaSpeicher2006} (Fig.~\ref{fig:intGraphs}b).
Some locality comes into play by considering $k$-local interactions --- each interaction $O_i$ acts on at most $k$ qudits. 
Mean-field models such as those of Sherrington-Kirkpatrick\cite{Sherrington1975} and Sachdev-Ye-Kitaev \cite{Sachdev1993,KitaevTalk2015,Kitaev2018}  are $k$-local in this sense despite all spins (qudits) interacting with one another (Fig.~\ref{fig:intGraphs}c). 
Most physical systems, however, are \emph{spatially} local wherein each degree of freedom only interacts with neighbors in a finite dimensional geometry (Fig.~\ref{fig:intGraphs}d).

The formalism we develop below encompasses all of these possible locality structures at large $N$.
It is well known that the $1/N$ expansion for correlators in random matrix theory may be viewed as a topological expansion in which the leading diagrams are planar\cite{Hooft1974,ZeeBook2010}.
This generalizes to arbitrary interaction graphs by the introduction of \emph{stacked} diagrams, which we define and analyze in Sec.~\ref{sec:planar}. 
The leading contributions at large-$N$ are \emph{stacked planar} and corrections arise as higher genus diagrams with multiple boundaries.

\begin{figure}[tbp]
\begin{center}
\includegraphics{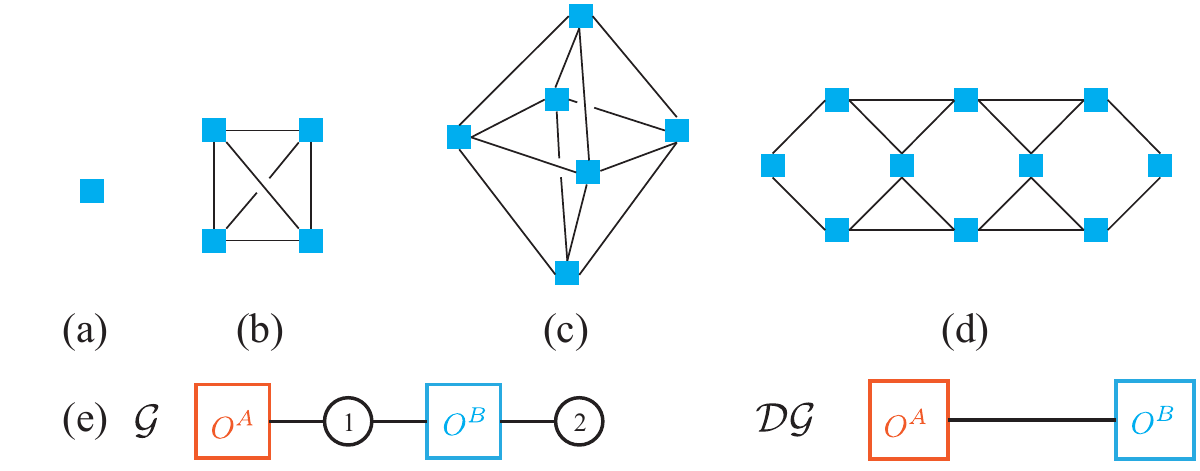}
\caption{(e) Left: The interaction graph $\mathcal{G}$ of a two qudit model with two interactions, $O^A$ and $O^B$. Circles denotes qudits and squares denote interactions. Right: Dependency graph $\mathcal{DG}$ corresponding to the interaction graph. Top: Dependency graphs corresponding to interaction graphs in Fig.~\ref{fig:intGraphs}.}
\label{fig:depGraphs}
\end{center}
\end{figure}

The stacked planar diagrams permit a crucial simplification at leading order: their structure and values depend only on the trace moments $\ldb (O^i)^m \rdb$ of the individual interaction terms and the \emph{dependency graph} $\mathcal{DG}$ of the interaction graph $\mathcal{G}$.  
The nodes of $\mathcal{DG}$ are the interactions (squares) in $\mathcal{G}$ with edges between two interactions whenever they share at least 1 qudit in $\mathcal{G}$~(Fig.~\ref{fig:depGraphs}). 
Systems with different interaction graphs can have the same dependency graph so this is a non-trivial reduction in the combinatorial data relevant to compute a correlator.
Moreover, this allows us to reformulate the diagrammatic sum as a sum over \emph{dependency partitions}, Eq.~\eqref{eq:dp_sum}, which we define.
Dependency partitions generalize the non-crossing partitions endemic to random matrix/free probability theory\cite{NicaSpeicher2006}.

From an algebraic point of view, the dependency graph $\mathcal{DG}$ encodes whether or not the operators $O^i$ commute because they act on distinct factors of the tensor product Hilbert space.
The dependency partitions allow crossings between operators which commute but are non-crossing between those which do not.
In Sec.~\ref{sec:heap_freeness}, we generalize the notion of free independence of non-commuting random variables to collections of variables which commute according to a given dependency graph $\mathcal{DG}$ and show that the $O^i$ of Eq.~\eqref{eq:generalham} are \emph{heap free} in this sense at leading order in large $N$.

The formulation in terms of heap free non-commuting random variables leads us in Sec.~\ref{sec:cumulantexp} to a  generalization of the moment-free cumulant expansion of free probability theory to heap free variables. 
This expansion, Eq.~\eqref{eq:free_cum_exp}, leads to the most compact combinatorial approach to evaluating the trace moments of Eq.~\eqref{eq:trace_moment_def}. 

Finally, as an application of these results, we show in Sec.~\ref{sec:qsat} that the Quantum Satisfiability problem reduces to the evaluation of the independence polynomial of $\mathcal{DG}$ in the large-N limit for a large class of graphs. 
This partially closes a conjecture of \cite{Sattath2016} regarding the tightness of the quantum Shearer bound.

There are several streams of prior work related to this manuscript.

The first prominent extension of random matrix theory to include spatial locality was Wegner's $n$-orbital model\cite{Wegner1979, BrezinZee1995}. 
However, Wegner's model can be considered as an extension of Anderson's model of disordered hopping\cite{Anderson1958} to $n$ levels and is thus still only a single particle problem. 
Recent work in the many-body context has focused on understanding the dynamics arising from time evolution under random unitary circuits\cite{Nahum2014, Keyserlingk2018, Khemani2017, Rakovszky2017, Chalker2017}. We hope that the tools developed in this paper will provide analytical control on the Hamiltonian version of these problems.

The Hamiltonian of Eq.~\eqref{eq:generalham} with all-to-all interactions (as in Fig.~\ref{fig:intGraphs}c) is closely related to a number of models of much recent interest in the study of quantum chaotic dynamics, holography and scrambling \cite{Hayden2007, Susskind2008, Roberts2015, Maldacena2016, Gharibyan2018}.
These are the embedded ensembles \cite{MonFrench1975}, Sachdev-Ye \cite{Sachdev1993} and Kitaev \cite{KitaevTalk2015,Kitaev2018} families of models describing randomly interacting fermions or bosons.
The connection is simplest in the case of $2$-local interactions.
Introducing an Abrikosov fermion $c_q$ with $N$ flavors on each qudit site $q$, $H$ becomes a quartic theory 
\begin{align}\label{eq:fermionic_twolocal}
    H&=-\sum J^{pq}_{\alpha \beta \gamma \delta} c^{\dagger \alpha}_p c^{\dagger \beta}_q c^\gamma_p c^\delta_q
\end{align}
Here, the $J^{qp}$ couplings are matrix elements of the interaction $O^{qp}$ between qudits $q$ and $p$. 
For Gaussian random interactions, $J^{qp}_{\alpha\beta\gamma\delta}$ is Gaussian with variance $\overline{J^2}\sim 1/N^2$ and mean $0$.
The Hamiltonian Eq.~\eqref{eq:fermionic_twolocal} is identical to those mentioned above up to symmetry. 
Eq.~\eqref{eq:fermionic_twolocal} has only a local $U(1)^n$ symmetry corresponding to the single conserved fermion on each site.
The Sachdev-Ye model has precisely the same structure but with an additional global $SU(N)$ symmetry as it arises from the fermion representation of interacting $SU(N)$ spins.
The embedded ensembles discard all of the symmetry except a global $U(1)$ corresponding to a fixed number of particles.
The Kitaev model possesses the least symmetry, having only a global $\mathbb{Z}_2$ corresponding to fermion parity%
\footnote{
For reference, the Hamiltonians, symmetries and number constraints for quartic fermionic variants of the various all-to-all models are:
\begin{align*}
    \begin{array}{lrlll}
    \textrm{Kitaev} & H_K &= -\sum J_{pqrs} \gamma_p \gamma_q \gamma_r \gamma_s & \mathbb{Z}_2 & \\
    \textrm{Embedded Ensemble} & H_{EE} &= -\sum J_{pqrs} c^\dagger_p c^\dagger_q c_r c_s & U(1) & \sum_q c^\dagger_q c_q = n_f\\
    \textrm{Sachdev-Ye} & H_{SY} &= -\sum J_{pq} c^{\dagger \mu}_p c^{\dagger \nu}_p c^\nu_p c^\mu_q & SU(N)\times U(1)^n &  \sum_\nu c^{\dagger\nu}_q c^{\nu}_q = n_f\\
    \textrm{Random Qudit} & H_{RQ} &= -\sum J^{pq}_{\alpha \beta \gamma \delta} c^{\dagger \alpha}_p c^{\dagger \beta}_q c^\gamma_p c^\delta_q  & U(1)^n &  \sum_\nu c^{\dagger \nu}_q c^\nu_q = 1
    \end{array}
\end{align*}
}.
We expect much of the physics of the these models to show up in Hamiltonian Eq.~\eqref{eq:fermionic_twolocal} at large $N$ and $n$.

Random matrices in the large-$N$ limit realize a non-commutative algebra described by free probability theory\cite{Voiculescu1991, NicaSpeicher2006}.
There has been recent mathematical progress in extending free probability to algebras with mixtures of classically independent and freely independent (non-commuting) variables.
This algebraic generalization has been dubbed both $\Lambda$-freeness\cite{Mlotkowski2004} and $\epsilon$-freeness\cite{Speicher2016}. 
From this perspective, the results of Sec.~\ref{sec:heap_freeness} show that random Hamiltonians of the type of Eq.~\eqref{eq:generalham} on extended interaction graphs provide a physically motivated realization of a $\Lambda$-free algebra at leading order in $N$.
Corrections to this limit can be described using the stacked  diagrams of Sec.~\ref{sec:planar}.

\section{Stacked planar diagrams and reduction to the dependency graph}
\label{sec:planar}

Each interaction $O^i$ can be represented in the $n$ qudit Hilbert space with a multi-index notation using $2n$ indices. For example, if $O^i$ acts on the first $k$ qudits,  
\begin{align*}
	O^i_{(\alpha_1 ... \alpha_n) (\beta_1 ... \beta_n)} &= 
	U^i_{ (\alpha_{1} ... \alpha_{k}) (\mu_1 ... \mu_k)} \lambda^{i}_{ (\mu_1 ... \mu_k) (\nu_1 ... \nu_k)} U^{i \dagger}_{(\nu_1 ... \nu_k) (\beta_{1} ... \beta_{k})} \delta_{(\alpha_{k+1} \cdots \alpha_n)( \beta_{k+1} \cdots \beta_n)}
\end{align*}
where a summation is implied over repeated indices. 
The non-trivial indices correspond to those qudits on which the interaction acts.
In the large $N$ limit, we take the diagonal matrices $\lambda^i$ to have a well-defined set of trace moments $\tr (\lambda^i)^p$ corresponding to an $O(1)$ spectrum. 

The multi-index notation lends itself naturally to a quantum circuit representation of moments. Consider a general operator product $O^{i_1} \cdots O^{i_p}$ built from the interactions in $H$. The Haar averaged trace moment
\begin{align}
\label{eq:general_op_moment}
	\left\ldb O^{i_1} \cdots O^{i_p} \right\rdb &\equiv \disavg{\tr (O^{i_1} \cdots O^{i_p})} 
	= \frac{1}{N^n} \disavg{\Tr ( U^{i_1} \lambda^{i_1} U^{i_1\dagger} \cdots U^{i_p} \lambda^{i_p} U^{i_p\dagger})}
\end{align}
may be viewed diagrammatically as the average of a periodic quantum circuit. For example, with the interaction graph given in Fig.~\ref{fig:depGraphs}a, 
\begin{align*}
  \ldb O^{A} O^{B} O^A O^B \rdb = \frac{1}{N^2}\disavg{\vcenter{\hbox{\includegraphics[scale=0.55]{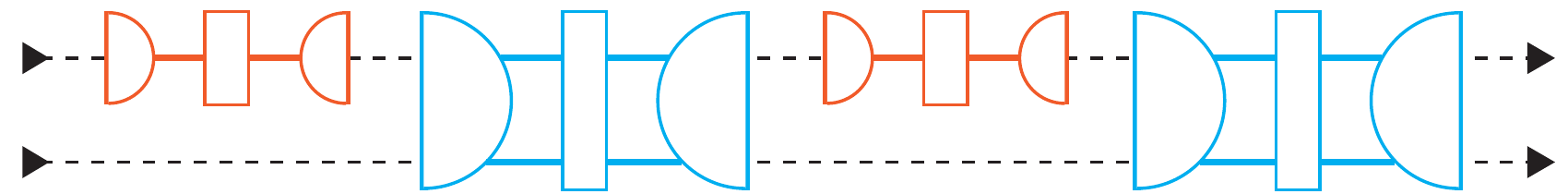}}}}_{U^A,U^B}
\end{align*}
Here, the semicircle gates represent $U^i$ and $U^{i \dagger}$ and the rectangles represent the diagonal matrix $\lambda^i$. 
We draw the lines connecting the unrotated basis of each qudit (outside conjugation by $U^i$) as dashed while those in the rotated basis (inside the conjugation by $U^i$) as solid.

\subsection{Diagrammatic averaging over unitary group}
\label{sub:diagrams_singlequdit}

To build a set of diagrammatic rules for dealing with the average over the $U^i$, let us first recall how to average over a single unitary $U \in U(N)$.

The key formula for averaging matrix elements of $U$ is
\begin{align}
\label{eq:weingarten_expansion}
	\disavg{U_{\alpha_1 \mu_1} U^\dagger_{\nu_1 \beta_1} \cdots U_{\alpha_p \mu_p} U^\dagger_{\nu_p \beta_p} } &=
	\sum_{\substack{\sigma, \tau \in S_p } }
	\Wg(\tau^{-1} \sigma; N) 
	\delta_{\alpha_1,\beta_{\sigma 1}} \cdots \delta_{\alpha_p,\beta_{\sigma p}} 
	\delta_{\mu_1,\nu_{\tau 1}} \cdots \delta_{\mu_p,\nu_{\tau p}} 
\end{align}
This formula provides an analog of Wick's theorem which allows us to express the Haar average as a sum over all pairings $\tau$ and $\sigma$ of inner $\mu, \nu$ (solid) and outer $\alpha, \beta$ (dashed) indices respectively from $U$ to $U^\dagger$. The coefficient $\Wg(\tau^{-1}\sigma; N)$ is known as the Weingarten function for the unitary group $U(N)$. See \cite{Weingarten1978, Collins2006} for more details. 
For our purposes, the most important features of $\Wg$ are that
\begin{enumerate}[(i)]
	\item $\Wg$ only depends on the conjugacy class of the permutation $\tau^{-1}\sigma$ within the permutation group $S_p$. That is, $\Wg(\tau^{-1}\sigma; N)$ only depends on the lengths of the cycles in the cycle decomposition of $\tau^{-1}\sigma$. This follows readily from the commutativity of the matrix elements on the left hand side of Eq.~\eqref{eq:weingarten_expansion}. 
	\item At leading order in large $N$, $\Wg$ in fact factorizes over the cycle decomposition,
	\begin{align}
		\label{eq:weingarten_largen}
		\Wg(\tau^{-1}\sigma; N) \asymp \prod_{\pi \in \textrm{Cycles}(\tau^{-1}\sigma)} \frac{ (-1)^{1+|\pi|} c_{|\pi|} }{N^{|\pi|-1}}
	\end{align}
	Here, $|\pi|$ is the length of the cycle $\pi$ and $c_{|\pi|}$ is the $|\pi|$'th Catalan number.
\end{enumerate}
We see that the \emph{relative} permutation between the dashed $\sigma$ and solid $\tau$ pairings is crucial for determining the $N$ dependence of the diagram corresponding to pairing $(\sigma, \tau)$.

Consider a single qudit of dimension $N$ with a single operator $O$ acting on it. 
Then $\ldb O^p \rdb = \ldb (U\lambda U^\dagger)^p \rdb = \tr \lambda^p$ is actually independent of $U$, but the diagrammatic expansion is non-trivial.
As a circuit, we can write
\begin{align}
	\ldb O^p \rdb &= \frac{1}{N} \disavg{ \vcenter{\hbox{\includegraphics{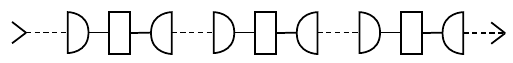}}}}_U
\end{align}
where we have drawn as an example the circuit for $p=3$.
To apply Eq.~\eqref{eq:weingarten_expansion}, we redraw this periodic circuit as a (counter-clockwise) oriented circle
\begin{align}
	\vcenter{\hbox{\includegraphics{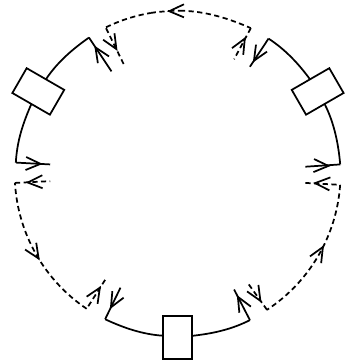}}}
\end{align}
where we replace the $U$ and $U^\dagger$ gates with \emph{boundary vertices} where each index line turns off the circular boundary and into the bulk of the circle, $\vcenter{\hbox{\includegraphics{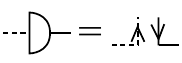}}},~ \vcenter{\hbox{\includegraphics{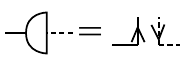}}}$.
The sum over $\sigma, \tau$ in Eq.~\eqref{eq:weingarten_expansion} corresponds to summing over all possible diagrams in which solid and dashed lines at each boundary vertex are paired across the bulk of the circle, consistent with the orienting arrows. 

Following t'Hooft \cite{Hooft1974, ArgamanZee1996, Brouwer1996}, it turns out that keeping track of the factors of $N$ associated to each diagram is greatly facilitated by using \emph{double line} or \emph{ribbon} notation, in which all bulk lines must run in pairs which do not separate. 
It is clear that this is possible when the relative permutation $\tau^{-1}\sigma=1$ is trivial, as this implies that the solid ($\tau$) and dashed ($\sigma$) lines pair the  indices of a given boundary $U$ to the same destination $U^\dagger$.
The nontrivial relative permutations are accommodated by introducing a degree $2k$ ``bulk vertex'' for each cycle of length $k$ in $\tau^{-1}\sigma$. For example, if $\tau^{-1} \sigma$ contains the cycle $(134)$, we draw
\begin{align}
	\vcenter{\hbox{\includegraphics{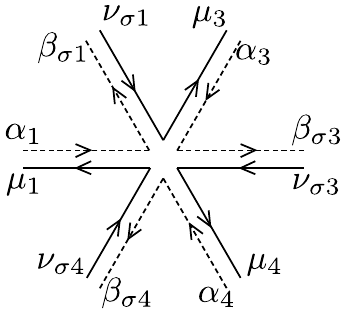}}}
\end{align}
which implements the relevant pairing without separating the double lines. Notice that every $\alpha$ pairs to the next $\beta$ by $\sigma$ while the $\nu$ pairs to $\mu$ by $\tau^{-1}$. 

Drawing the diagrams using these rules allows us to interpret each diagram in $\ldb O^p \rdb$ as an oriented surface with boundary given by the circle. The value of such a diagram is then given by a product of
\begin{align*}
\begin{array}{ll}
	N (-1)^{k} c_k  &\textrm{for each degree } 2k \textrm{ vertex} \\
	N^{-1} &\textrm{for each ribbon} \\
	N & \textrm{for each dashed loop} \\
	N \tr (\lambda^l) & \textrm{for each solid loop going through $l$ boundary squares}	
\end{array}
\end{align*}
The total $N$ dependence is thus
\begin{align}
	N^{f - e_{\textrm{bulk}} + v_{\textrm{bulk}}} = N^{f - (e_{\textrm{bulk}} + e_{\textrm{boundary}}) + (v_{\textrm{bulk}} + v_{\textrm{boundary}})} = N^\chi
\end{align}
where $\chi$ is the Euler characteristic of the surface, $f$ is the number of faces (closed loops) and $e_{\textrm{bulk}}, e_{\textrm{boundary}}$ and $v_{\textrm{bulk}}, v_{\textrm{boundary}}$ are the number of bulk and boundary edges and vertices respectively. We have used the fact that $e_{\textrm{boundary}} = v_{\textrm{boundary}} = 2p$. 
For a surface with a circular boundary, the Euler characteristic $\chi \le 1$ with the maximum obtained for topological discs -- that is, planar diagrams.

For example,
\begin{align}
	\ldb O \rdb = \frac{1}{N} \vcenter{\hbox{\includegraphics{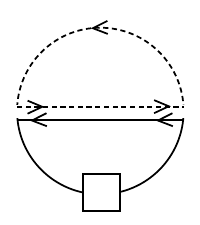}}} = \frac{1}{N} \Tr \lambda = \tr \lambda
\end{align}
consists of a single planar diagram. While,
\begin{align}
	\ldb O^2 \rdb &= \frac{1}{N} \left( 
	\vcenter{\hbox{\includegraphics{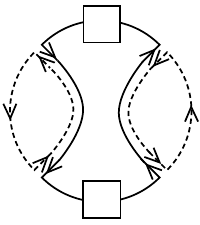}}} 
	+ \vcenter{\hbox{\includegraphics{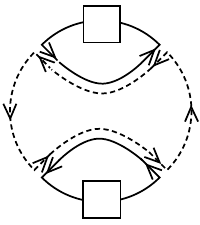}}} 
	+ \vcenter{\hbox{\includegraphics{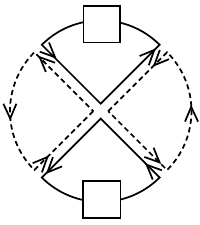}}} 
	+ \vcenter{\hbox{\includegraphics{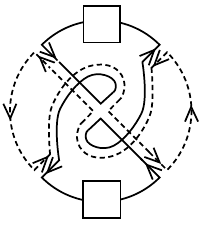}}} 
	\right) \\
	&=  \tr \lambda^2 + (\tr \lambda)^2 - (\tr \lambda)^2 - \frac{1}{N^2} \tr \lambda^2  \xrightarrow[N\to\infty]{} \tr \lambda^2 \nonumber
\end{align}
consists of three planar diagrams at leading order and one suppressed non-planar 
diagram\footnote{We note that the sub-leading in $N$ contributions do not cancel order by order unless one keeps next order corrections to the Weingarten function.}.

Generalizing these diagrammatics to multiple independent interactions acting on the same qudit is straightforward. 
The Haar average of the matrix elements factors across $U^i, U^{i\dagger}$ with different labels $i$, so one need simply sum over all pairings with the same rules as above but which additionally respect the label $i$ of each boundary vertex. 
This is represented in our figures by the \emph{color} of the boundary vertices and corresponding bulk ribbons. 
There are accordingly $M$ colors labeled by $i$ for moments arising from an interaction graph with $M$ operators $O^i$.
The $N$ dependence of such a colored diagram is still given by its Euler characteristic and only planar diagrams contribute to leading order as $N\to\infty$.
In other words, only the \emph{monochromatic} subset of planar diagrams, in which each solid loop is a single color, contribute.

\subsection{Stacked planar diagrams in the multi-qudit case}
\label{sub:stackedplanar}

We now return to the multiqudit case. 
For each interaction of type $i$, the indices $\alpha, \beta, \mu, \nu$ attached to $U^i_{\alpha,\mu}, U^{i \dagger}_{\beta,\nu}$ in Eq.~\eqref{eq:weingarten_expansion} should now be interpreted as multi-indices $\alpha = (\alpha^1 \alpha^2 \cdots \alpha^{|\partial i|})$ corresponding to the $|\partial i|$ qudits on which $O^i$ acts. 
Since,
\begin{align}
	\delta_{\alpha \beta} = \delta_{\alpha^1\beta^1}\delta_{\alpha^2 \beta^2}\cdots\delta_{\alpha^{|\partial i|}\beta^{|\partial i|}}
\end{align}
factors, the pairings $\sigma, \tau$ may be viewed as connecting the lines on each of the $|\partial i|$ layers in parallel. 
For a given interaction $O^i$, the local Hilbert space dimension also factors across layers. 
Thus, we have the following rules:
\begin{enumerate}
	\item Draw a stack of $n$ boundary circles (corresponding to each qudit) from the circuit $\Tr O^{i_1} \cdots O^{i_p}$. Note the boundary circle on layer $q$ only has boundary gates $\lambda^i$ and boundary vertices for the $O^i$ which act on that layer.

	\item For each type of interaction $O^i$ (color), draw all possible bulk pairings with the ribbon and bulk vertex rules \emph{locked} together across the relevant layers, i.e., choosing a pairing on one layer repeats the same pattern on all layers on which the interaction acts.
\end{enumerate}
If the diagram has Euler characteristic $\chi_q$ on qudit layer $q$, then the total $N$ dependence is given by $\prod_q N^{\chi_q}$.
The leading diagrams are thus stacks of planar discs with $\chi_q = 1$ for all layers $q$. 

\begin{figure}[tbp]
\begin{center}
\includegraphics[width=\textwidth,height=\textheight,keepaspectratio]{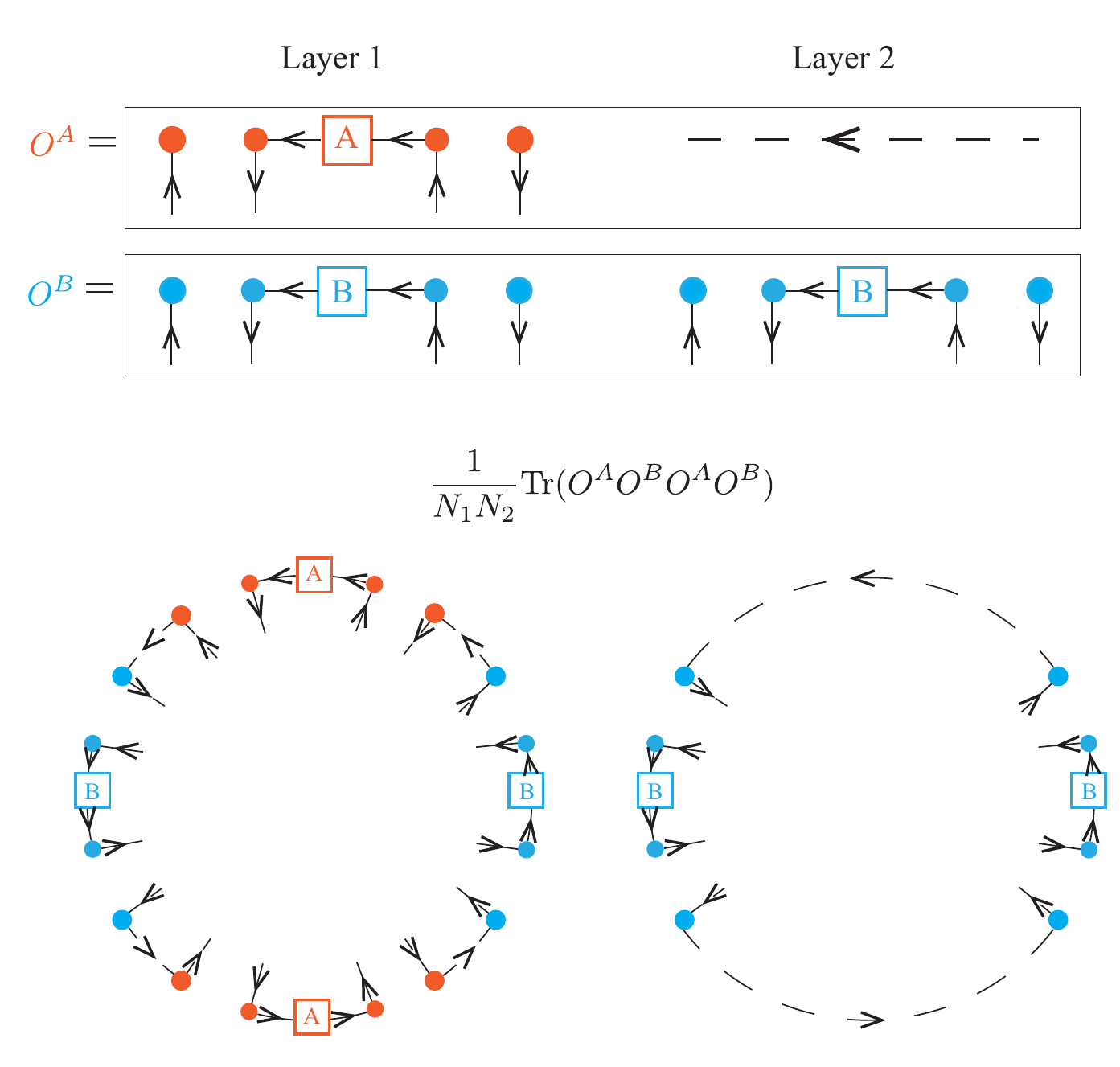}
\caption{How to construct the diagrammatic representation of $\ldb O^A O^B O^A O^B \rdb$ for the interaction graph of Fig.~\ref{fig:depGraphs}e. 
$N_1$ and $N_2$ denote the local Hilbert space of qudit 1 and 2 respectively.
(top) The piece of boundary circle corresponding to $O^A$ and $O^B$ on layers $1$ and $2$ (left and right) of the stacked diagram. Here, the gate containing $A$ ($B$) corresponds to the diagonal form of $O^A$ ($O^B$).
(bottom) The boundary circles for representing the correlator $\ldb O^A O^B O^A O^B \rdb$ on layers $1$ and $2$ (left and right) prior to summing over bulk pairings which occurs when doing the disorder average.} 
\label{fig:diagRep}
\end{center}
\end{figure}

\begin{figure}[htbp]
\begin{center}
\includegraphics[width=\textwidth,height=0.92\textheight,keepaspectratio]{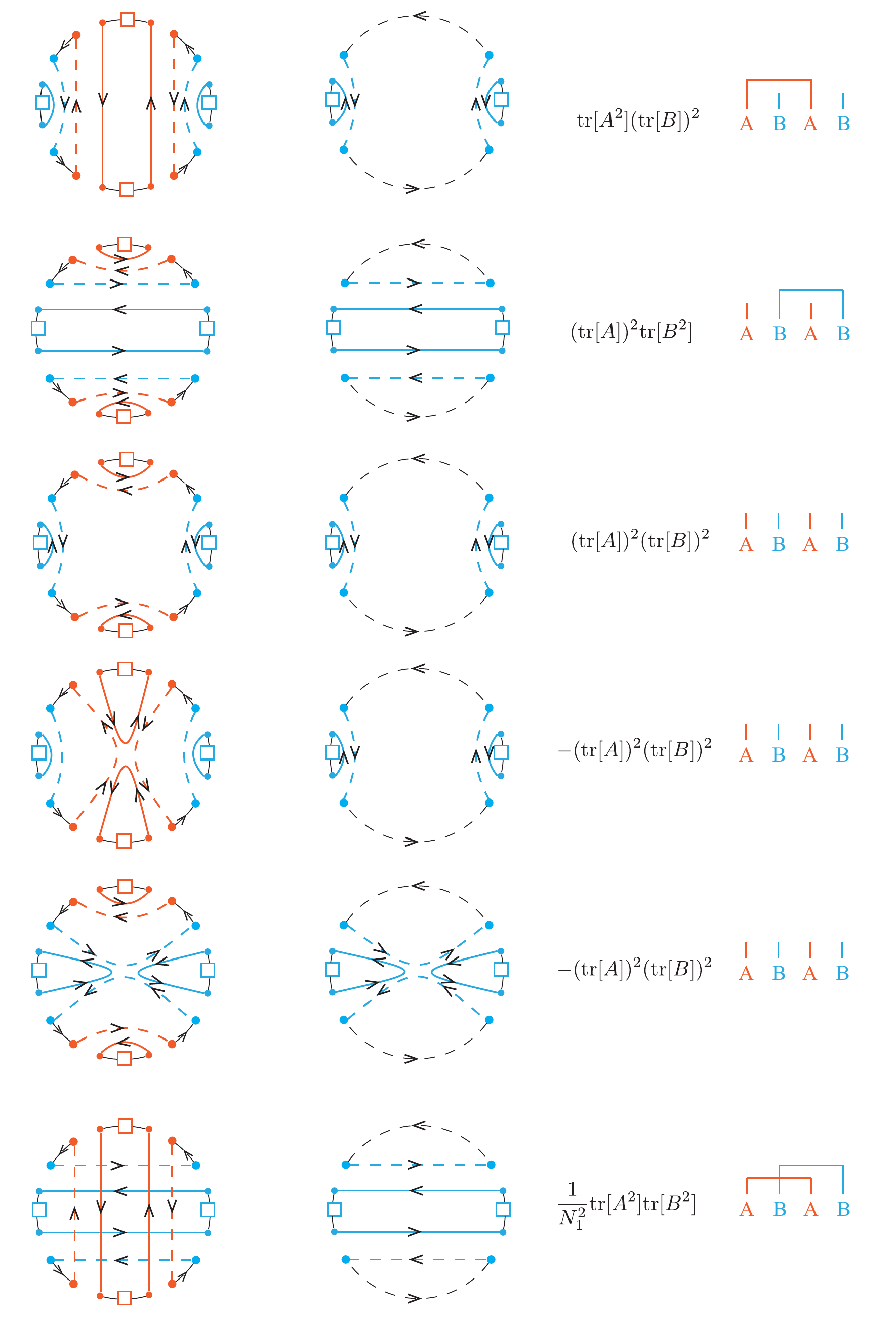}
\caption{
Some of the stacked diagrams contributing to $\ldb O_A O_B O_A O_B \rdb \equiv \frac{1}{N_1 N_2}\mathbb{E}[\mathrm{Tr}(O_A O_B O_A O_B)]$ for the model of Fig.~\ref{fig:diagRep}. Only stacked planar diagrams (first 5) contribute at leading order in large $N_1$. Although we usually take all $N_q=N$, this example also shows that not all qudit dimensions need to be taken to be large to achieve the planar reduction. The associated partitions are illustrated on the right. The dependency partitions here would exclude the last crossing partition.}
\label{fig:ABABdiagrams}
\end{center}
\end{figure}

Figures~\ref{fig:diagRep} and \ref{fig:ABABdiagrams} illustrate how to apply the above recipe to construct the complete set of monochromatic stacked planar diagrams contributing to $\ldb O^A O^B O^A O^B \rdb$ for a simple two qudit interaction graph. 

The full value of a stacked diagram is not quite a product over layers. Rather, one obtains 
\begin{align*}
	\begin{array}{ll} 
		(-1)^{k+1}c_k & \textrm{for each } 2k \textrm{ bulk vertex counting locked layers once} \\
		N & \textrm{for each bulk vertex on each layer }\\
		N^{-1} & \textrm{for each bulk ribbon on each layer } \\
		N & \textrm{for each dashed loop on each layer }\\
		N^{|\partial i|}\tr (O^i)^l & \textrm{for each solid loop of type $i$ counting locked layers once} 
	\end{array}
\end{align*}

There is one important generalization of the stacked diagrams: 
since $P(O^i) = P(U^i \lambda^i U^{i \dagger}) = U^i P(\lambda^i) U^{i \dagger}$ for any polynomial $P$, the stacked diagram expansion can equally well be applied to moments of the form $\ldb P^{i_1}(O^{i_1}) \cdots P^{i_p}(O^{i_p}) \rdb$, with the boundary gates representing $P^{i_j}(\lambda^{i_j})$ carrying the color $i_j$.

To summarize, at leading order in the large $N$ limit, disorder averaged trace moments are given by a sum over monochromatic stacked planar diagrams
\begin{align}
\label{eq:mono_stacked_planar}
    \ldb P^{i_1}(O^{i_1}) \cdots P^{i_p}(O^{i_p}) \rdb &= \sum_{\substack{\textrm{monochromatic}\\\textrm{stacked planar} \\ \textrm{diagrams}}} \left( \prod_{\substack{\textrm{degree } 2k \\ \textrm{ vertices}}} (-1)^{k+1}c_k \right) \left( \prod_{\substack{\textrm{solid}\\\textrm{loops } l}} \tr \left(P^{l_1}(O^{l_1})  \cdots P^{l_{|l|}}(O^{l_{|l|}})\right) \right)
\end{align}
Though there are far fewer stacked planar diagrams than the total number of terms in the Weingarten expansion, the number of diagrams grows  rapidly with $p$ (and depends in detail on $\mathcal{DG}$) and, worse for high order calculations, they come with varying signs. 
Nonetheless, this representation will allow us to derive several more efficient representations below.

\subsection{Dependency partitions}
\label{sub:dependencypartitions}
 
The $1/N$ expansion can in principle be organized as a topological expansion of the stacked diagrams. 
However, even the leading stacked planar diagrams are rather complicated to enumerate and analyze.
In this section, we will reorganize the stacked \emph{planar} diagrams in terms of certain partitions of the string of $p$ operators $O^{i_1} \cdots O^{i_p}$. 
The partitions are both easier to visualize and encode the data contributing to the trace moments more compactly.
Before we can state the result, we need a few definitions regarding partitions.

A \emph{monochromatic partition} of a string of $p$ operators $O^{i_1} \cdots O^{i_p}$ is a decomposition of the operators into disjoint blocks $B$ such that each block contains only operators of one color. 
Such partitions are conveniently represented by connections drawn above the string indicating the blocks.
For example, the monochromatic partitions of $ABAB$ are (using the shorthand $A=O^A$, $B=O^B$),
\tikzstyle{srp}+=[baseline,remember picture]
\tikzstyle{spart}=[anchor=base,use as bounding box, inner sep=0pt, outer sep=0pt]
\newcommand{\ABAB}{\tikz[srp]{\node[spart] (A1) {\text{$A$}}}
    \tikz[srp]{\node[spart] (B1) {\text{$B$}}}
    \tikz[srp]{\node[spart] (A2) {\text{$A$}}}
    \tikz[srp]{\node[spart] (B2) {\text{$B$}}}}
\begin{align*}
    \ABAB
    \begin{tikzpicture}[srp,overlay]
        \draw [thick,color=red] ($(A1.north)+(0,3pt)$) -- +(0,4pt);
        \draw [thick,color=blue] ($(B1.north)+(0,3pt)$) -- +(0,4pt);
        \draw [thick,color=red] ($(A2.north)+(0,3pt)$) -- +(0,4pt);
        \draw [thick,color=blue] ($(B2.north)+(0,3pt)$) -- +(0,4pt);
    \end{tikzpicture}
    \qquad
    \ABAB
    \begin{tikzpicture}[srp,overlay]
        \draw [thick,color=red] ($(A1.north)+(0,3pt)$) -- +(0,8pt) -| ($(A2.north)+(0,3pt)$);
        \draw [thick,color=blue] ($(B1.north)+(0,3pt)$) -- +(0,4pt);
        \draw [thick,color=blue] ($(B2.north)+(0,3pt)$) -- +(0,4pt);
    \end{tikzpicture}
    \qquad
    \ABAB
    \begin{tikzpicture}[srp,overlay]
        \draw [thick,color=blue] ($(B1.north)+(0,3pt)$) -- +(0,8pt) -| ($(B2.north)+(0,3pt)$);
        \draw [thick,color=red] ($(A1.north)+(0,3pt)$) -- +(0,4pt);
        \draw [thick,color=red] ($(A2.north)+(0,3pt)$) -- +(0,4pt);
    \end{tikzpicture}
    \qquad
    \ABAB   \begin{tikzpicture}[srp,overlay]
        \draw [thick,color=red] ($(A1.north)+(0,3pt)$) -- +(0,4pt) -| ($(A2.north)+(0,3pt)$);
        \draw [thick,color=blue] ($(B1.north)+(0,3pt)$) -- +(0,8pt) -| ($(B2.north)+(0,3pt)$);
    \end{tikzpicture}
    \vbox to 20pt{} 
\end{align*}
The first partition has four length 1 blocks; the second has one length two and two length one, etc. The five partitions of $AAA$ are
\newcommand{\AAA}{\tikz[srp]{\node[spart] (A1) {\text{$A$}}}
    \tikz[srp]{\node[spart] (A2) {\text{$A$}}}
    \tikz[srp]{\node[spart] (A3) {\text{$A$}}}}
\begin{align*}
    \AAA
    \begin{tikzpicture}[srp,overlay]
        \draw [thick,color=red] ($(A1.north)+(0,3pt)$) -- +(0,4pt);
        \draw [thick,color=red] ($(A2.north)+(0,3pt)$) -- +(0,4pt);
        \draw [thick,color=red] ($(A3.north)+(0,3pt)$) -- +(0,4pt);
    \end{tikzpicture}
    \qquad
    \AAA
    \begin{tikzpicture}[srp,overlay]
        \draw [thick,color=red] ($(A1.north)+(0,3pt)$) -- +(0,4pt) -| ($(A2.north)+(0,3pt)$);
        \draw [thick,color=red] ($(A3.north)+(0,3pt)$) -- +(0,4pt);
    \end{tikzpicture}
    \qquad
    \AAA
    \begin{tikzpicture}[srp,overlay]
        \draw [thick,color=red] ($(A1.north)+(0,3pt)$) -- +(0,8pt) -| ($(A3.north)+(0,3pt)$);
        \draw [thick,color=red] ($(A2.north)+(0,3pt)$) -- +(0,4pt);
    \end{tikzpicture}
    \qquad
    \AAA
    \begin{tikzpicture}[srp,overlay]
        \draw [thick,color=red] ($(A2.north)+(0,3pt)$) -- +(0,4pt) -| ($(A3.north)+(0,3pt)$);
        \draw [thick,color=red] ($(A1.north)+(0,3pt)$) -- +(0,4pt);
    \end{tikzpicture}
    \qquad
    \AAA
    \begin{tikzpicture}[srp,overlay]
        \draw [thick,color=red] ($(A1.north)+(0,3pt)$) -- +(0,4pt) -| ($(A2.north)+(0,3pt)$);
        \draw [thick,color=red] ($(A2.north)+(0,7pt)$) -|
        ($(A3.north)+(0,3pt)$);
    \end{tikzpicture}
    \vbox to 15pt{}
\end{align*}
where the last partition illustrates a block of length 3.
\emph{Non-crossing} partitions are those whose connection diagram may be drawn without any crossings. For example, the monochromatic non-crossing partitions of $ABAB$ are
\begin{align*}
    \ABAB
    \begin{tikzpicture}[srp,overlay]
        \draw [thick,color=red] ($(A1.north)+(0,3pt)$) -- +(0,4pt);
        \draw [thick,color=blue] ($(B1.north)+(0,3pt)$) -- +(0,4pt);
        \draw [thick,color=red] ($(A2.north)+(0,3pt)$) -- +(0,4pt);
        \draw [thick,color=blue] ($(B2.north)+(0,3pt)$) -- +(0,4pt);
    \end{tikzpicture}
    \qquad
    \ABAB
    \begin{tikzpicture}[srp,overlay]
        \draw [thick,color=red] ($(A1.north)+(0,3pt)$) -- +(0,8pt) -| ($(A2.north)+(0,3pt)$);
        \draw [thick,color=blue] ($(B1.north)+(0,3pt)$) -- +(0,4pt);
        \draw [thick,color=blue] ($(B2.north)+(0,3pt)$) -- +(0,4pt);
    \end{tikzpicture}
    \qquad
    \ABAB
    \begin{tikzpicture}[srp,overlay]
        \draw [thick,color=blue] ($(B1.north)+(0,3pt)$) -- +(0,8pt) -| ($(B2.north)+(0,3pt)$);
        \draw [thick,color=red] ($(A1.north)+(0,3pt)$) -- +(0,4pt);
        \draw [thick,color=red] ($(A2.north)+(0,3pt)$) -- +(0,4pt);
    \end{tikzpicture}
    \vbox to 20pt{}
\end{align*}

Let us return briefly to the case where there is only one qudit degree of freedom on which the operator $O$ acts, as discussed in Sec.~\ref{sub:diagrams_singlequdit}. 
The moments $\ldb O^p \rdb$ can be expanded as a sum of single layer planar diagrams with $p$ boundary operators $O$.
To each planar diagram, we may associate a non-crossing partition $\tau$ of the $p$ operators by grouping them into blocks according to the solid loops which connect them\footnote{Formally, since the solid loops are given by the cycles of the permutation $\tau$ representing the $\mu$-index pairing, the blocks $B$ in the partition are given by the cycle decomposition of $\tau$. In fact, since the solid loops defined by $\tau$ are planar, the cycles must be order preserving (ie. $(135)$ is an allowed cycle but not $(153)$). Thus, planar $\tau$ are actually in one-to-one correspondence with their cycle decompositions.  We slightly abuse notation by using $\tau$ to represent both the $\mu$-index pairing and the corresponding partition.}.
Planarity implies that the solid faces cannot overlap -- and that the relevant partitions are likewise non-crossing.
Regrouping the sum over diagrams by partition $\tau$, we are led to the representation
\begin{align}
    \ldb O^p \rdb &= \sum_{\tau \in \NC(p)} C_\tau \prod_{B \in \tau} \tr O^{|B|}
\end{align}
where $\NC(p)$ indicates the set of non-crossing partitions of $p$ objects and 
\begin{align}
    C_\tau &= \sum_{\sigma | (\sigma,\tau) \textrm{ is planar}} \left( \prod_{\substack{\textrm{degree }2k\textrm{ vertices}\\\textrm{in }(\sigma,\tau)}} (-1)^{k+1} c_k\right)
\end{align}
accumulates the Catalan coefficients coming from each planar diagram $(\sigma, \tau)$ with the same solid partition $\tau$.

If multiple $O^i$ act on the same single qudit (as in Fig.~\ref{fig:intGraphs}b), the relevant partitions $\tau$ are both non-crossing and monochromatic, just like the solid loops in the single layer diagrams. 

Returning to the general case, the sum over stacked planar diagrams can be regrouped into a sum over \emph{monochromatic dependency partitions} of the string of operators,
\begin{align}
\label{eq:dp_sum}
    \ldb O^{i_1}\cdots O^{i_p} \rdb &= \sum_{\tau \in \DP} C_{\tau} \prod_{B\in \tau} \tr O^{B_1} \cdots O^{B_{|B|}}.
\end{align}
or, again generalizing $O^i$ to $P^i(O^i)$ as in Eq.~\eqref{eq:mono_stacked_planar},
\begin{align}
\label{eq:dp_sum_poly}
    \ldb P^{i_1}(O^{i_1})\cdots P^{i_p}(O^{i_p}) \rdb &= \sum_{\tau \in \DP} C_{\tau} \prod_{B\in \tau} \tr P^{B_1}(O^{B_1}) \cdots P^{B_{|B|}}(O^{B_{|B|}}).
\end{align}
The dependency partitions $\DP$ of a given string of $p$ operators must be 
\begin{itemize}
    \item monochromatic -- each block only connects operators of the same color, and,
    \item non-crossing between dependent colors (and between blocks of the same color).
\end{itemize}
For example, if the operators $A$, $B$, $C$ have the dependency graph
\begin{align}
\label{eq:abc_depgraph}
    \begin{tikzpicture}[inner sep=1mm,outer sep=0mm]
        \node [draw,red] (A) at (0,0) {$A$};
        \node [draw,blue] (B) at (1,0) {$B$};
        \node [draw,green] (C) at (2,0) {$C$};
        \draw [thick, black] (A) -- (B) -- (C);
    \end{tikzpicture}
\end{align}
then the monochromatic dependency partitions of $ABCABC$ are
\newcommand{\ABCABC}{%
    \tikz[srp]{\node[spart] (A1) {\text{$A$}}}
    \tikz[srp]{\node[spart] (B1) {\text{$B$}}}
    \tikz[srp]{\node[spart] (C1) {\text{$C$}}}
    \tikz[srp]{\node[spart] (A2) {\text{$A$}}}
    \tikz[srp]{\node[spart] (B2) {\text{$B$}}}
    \tikz[srp]{\node[spart] (C2) {\text{$C$}}}
    }
\begin{align*}
    \ABCABC
    \begin{tikzpicture}[srp,overlay]
        \draw [thick,red] ($(A1.north)+(0,3pt)$) -- +(0,4pt);
        \draw [thick,red] ($(A2.north)+(0,3pt)$) -- +(0,4pt);
        \draw [thick,blue] ($(B1.north)+(0,3pt)$) -- +(0,4pt);
        \draw [thick,blue] ($(B2.north)+(0,3pt)$) -- +(0,4pt);
        \draw [thick,green] ($(C1.north)+(0,3pt)$) -- +(0,4pt);
        \draw [thick,green] ($(C2.north)+(0,3pt)$) -- +(0,4pt);
    \end{tikzpicture}
    \qquad
    \ABCABC
    \begin{tikzpicture}[srp,overlay]
        \draw [thick,red] ($(A1.north)+(0,3pt)$) -- +(0,8pt) -|  ($(A2.north)+(0,3pt)$);
        \draw [thick,blue] ($(B1.north)+(0,3pt)$) -- +(0,4pt);
        \draw [thick,blue] ($(B2.north)+(0,3pt)$) -- +(0,4pt);
        \draw [thick,green] ($(C1.north)+(0,3pt)$) -- +(0,4pt);
        \draw [thick,green] ($(C2.north)+(0,3pt)$) -- +(0,4pt);
    \end{tikzpicture}
    \qquad
    \ABCABC
    \begin{tikzpicture}[srp,overlay]
        \draw [thick,red] ($(A1.north)+(0,3pt)$) -- +(0,4pt);
        \draw [thick,red] ($(A2.north)+(0,3pt)$) -- +(0,4pt);
        \draw [thick,blue] ($(B1.north)+(0,3pt)$) -- +(0,8pt) -| ($(B2.north)+(0,3pt)$);
        \draw [thick,green] ($(C1.north)+(0,3pt)$) -- +(0,4pt);
        \draw [thick,green] ($(C2.north)+(0,3pt)$) -- +(0,4pt);
    \end{tikzpicture}
    \qquad
    \ABCABC
    \begin{tikzpicture}[srp,overlay]
        \draw [thick,red] ($(A1.north)+(0,3pt)$) -- +(0,4pt);
        \draw [thick,red] ($(A2.north)+(0,3pt)$) -- +(0,4pt);
        \draw [thick,blue] ($(B1.north)+(0,3pt)$) -- +(0,4pt);
        \draw [thick,blue] ($(B2.north)+(0,3pt)$) -- +(0,4pt);
        \draw [thick,green] ($(C1.north)+(0,3pt)$) -- +(0,8pt) -| ($(C2.north)+(0,3pt)$);
    \end{tikzpicture}
    \qquad
    \ABCABC
    \begin{tikzpicture}[srp,overlay]
        \draw [thick,red] ($(A1.north)+(0,3pt)$) -- +(0,8pt) -| ($(A2.north)+(0,3pt)$);
        \draw [thick,blue] ($(B1.north)+(0,3pt)$) -- +(0,4pt);
        \draw [thick,blue] ($(B2.north)+(0,3pt)$) -- +(0,4pt);
        \draw [thick,green] ($(C1.north)+(0,3pt)$) -- +(0,12pt) -| ($(C2.north)+(0,3pt)$);
    \end{tikzpicture}
    \vbox to 20pt{}
\end{align*}
The last partition is allowed because $A$ and $C$ are not connected in $\mathcal{DG}$. Crossings between $A$ and $B$ blocks or $B$ and $C$ blocks are disallowed.

These rules follow directly from stacked planarity -- if two operators are dependent, they act on a shared planar layer and the corresponding solid faces cannot cross. 
On the other hand, the solid faces associated with operators which do not act on a shared qudit (ie. are not connected in $\mathcal{DG}$) have no such non-crossing restriction.
Given a stacked planar diagram, one can read off the corresponding partition $\tau$ by  `looking down' at the diagram from above and drawing the skeleton of the solid faces, recalling that we need only draw the skeleton for one copy of each solid loop across the stacked layers. 

As an example, the five stacked planar diagrams contributing to $\ldb O^A O^B O^A O^B \rdb$ in Fig.~\ref{fig:ABABdiagrams} can be regrouped according to their partitions as shown in the rightmost column. 

The regrouping of the stacked planar diagram sum in \eqref{eq:dp_sum} is of limited explicit calculational utility since the coefficients $C_\tau$ are still rather complicated.
However, it allows several general properties to be proven readily as we will show below and in the following sections.

For example, the form of Eq.~\eqref{eq:dp_sum} makes clear that the correlators $\ldb O^{i_1} \cdots O^{i_p} \rdb$ depend only on the dependency graph $\mathcal{DG}$ of the interaction graph $\mathcal{G}$, and the  moments $\tr (O^{i})^p$ of the individual operators.
This follows because the dependency partitions $\tau$ are determined by the non-crossing of \emph{dependent} colors, as are the dashed line pairings $\sigma$ which contribute to the coefficients $C_\tau$.
In particular, whenever the interactions $O^i$ have a fully connected dependency graph $\mathcal{DG}$, the correlators reduce precisely to those of a collection of interactions acting on a single qudit even if the underlying interaction graph $\mathcal{G}$ is more complicated. 

\section{Heap freeness}
\label{sec:heap_freeness}

Random matrices in the large-N limit realize a non-commutative algebra described by free probability theory~\cite{Voiculescu1991, NicaSpeicher2006}.
This abstraction provides a convenient way to compute moments directly in the large-N limit without explicitly summing diagrams. 
The central definition is that of free independence:
a collection of non-commuting operators $\{ O^i \}$ is \emph{free} or \emph{freely independent} if all alternating centralized moments involving them are zero. That is, for all polynomials $P_j$,
\begin{align}
\label{eq:cent_mom_zero}
    \Big\ldb \left(P_1(O^{i_1}) - \ldb P_1(O^{i_1}) \rdb \right)
    \left(P_2(O^{i_2}) - \ldb P_2(O^{i_2}) \rdb \right)
    \cdots 
    \left(P_p(O^{i_p}) - \ldb P_n(O^{i_p}) \rdb \right)
    \Big\rdb =0
\end{align}
where the operators `alternate', $O^{i_j} \ne O^{i_{j+1}}$, for $j=1\cdots p$.
From this property and linearity, \emph{all} mixed moments of products involving $\{ O^i \}$ are determined in terms of their individual moments $\ldb (O^i)^p \rdb$ recursively.
For example, one can calculate $\ldb O^1 O^2 O^1 O^2 \rdb = \ldb (O^1)^2 \rdb \ldb O^2 \rdb^2 + \ldb O^1 \rdb^2 \ldb (O^2)^2 \rdb - \ldb O^1 \rdb^2 \ldb O^2 \rdb^2$
using Eq.~\eqref{eq:cent_mom_zero} recursively.
It is well-known that a collection of independently Haar spun random matrices -- for us, a collection of interactions $O^i$ which act on the same qudit -- become freely independent in the large-N limit.

The usual independence of commuting random variables can also be expressed in terms of certain centralized moments vanishing. 
A collection of commuting variables $\{ O^i \}$ is independent if and only if
Eq.~\eqref{eq:cent_mom_zero} holds for all polynomials $P_j$ where each $i_j$ is distinct from all of the others.
Since the variables commute, this property  determines all of the mixed moments recursively.
This condition characterizes the moments of collections of interactions $O^i$ which act on disjoint qudits in $\mathcal{G}$, since they commute (at any $N$).

\subsection{Heaps}
\label{sub:heaps}

In general, the interactions $O^i$ have a dependency graph which is neither fully connected nor fully disconnected.
Accordingly, the operators are neither freely nor classically independent in the large $N$ limit.
Nonetheless, as we will shortly see, the vanishing of a certain class of centralized moments still holds: centralized moments of \emph{heaps}. 

\begin{figure}[tbp]
\begin{center}
\includegraphics[width=0.5\textwidth]{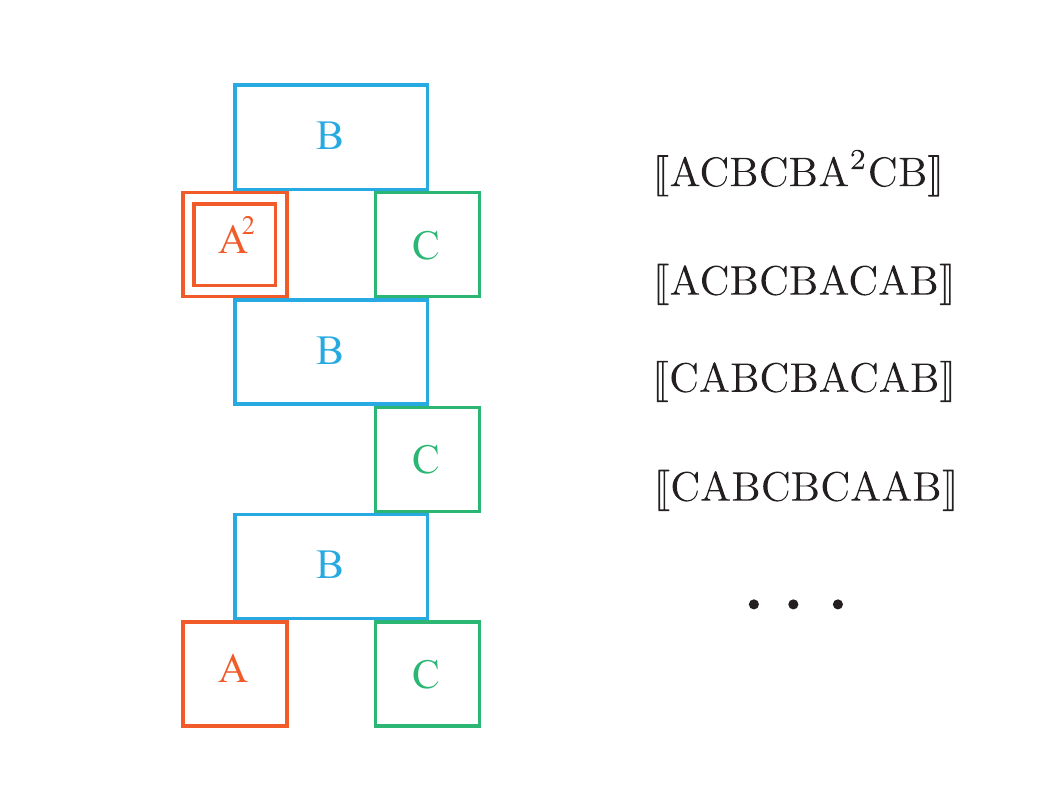}
\caption{Heap representation of a word with the canonical form $\llbracket \mathrm{A} \mathrm{C} \mathrm{B} \mathrm{C} \mathrm{B} \mathrm{A}^2 \mathrm{C} \mathrm{B} \rrbracket$ where A, B and C have the dependency graph of an open chain, Eq.~\eqref{eq:abc_depgraph}. 
The expressions on the right are all equivalent and produce the same heap on the left.
}
\label{fig:heap}
\end{center}
\end{figure}

Heaps provide a canonical representation of operator strings $O^{i_1} \cdots O^{i_p}$ where some of the operators commute%
\footnote{Formally, heaps provide a canonical form for the word problem on freely generated monoids whose only relation is that of commutation of certain generators.\cite{Viennot1989, Krattenthaler2006}}\cite{Viennot1989, Krattenthaler2006}.
Given interactions $\{O_{i}\}$ with dependency graph $\textbf{D}\mathcal{G}$, 
we construct a heap graphically by the following rules 
\begin{enumerate}
\item View each operator as a brick which overlaps with bricks of nearest-neighbor operators on $\textbf{D}\mathcal{G}$. 
\item A string of operators  $O^{i_1}O^{i_2}...$ is represented by dropping the bricks corresponding to each operator in order into a pile\footnote{Cf. Tetris.}. 
\item When bricks overlap, they stack on top of each other to form multiple rows. 
\item When a brick $O^i$ falls on top of another brick representing $(O^i)^m$, it merges downward to form a single brick representing $(O^i)^{m+1}$. 
\end{enumerate}
After all this stacking, we write a canonical form for the operator string by reading the bricks $W^j$ from bottom to top, with a fixed ordering of the bricks within a given row, $W^1 W^2 \cdots W^{p'}$. 
An example is shown in Fig.~\ref{fig:heap}.

Before moving on, let us point out several important properties of the heap form of an operator string, $W^1 W^2 \cdots W^{p'}$. 
Each $W^j = (O^i)^m$ is a monomial built out of a single color of interaction operator $O^i$. 
We may generalize heaps immediately to allow $W^j$ to be an arbitrary non-constant polynomial in the  operator $O^i$.
If two bricks $W^{j}$, $W^{l}$, with $j < l$, are of the same color, then there must be a dependent brick $W^k$  between them ($j < k < l$), holding them apart in the pile (adjacent bricks cannot be the same color). 

\subsection{Vanishing of centralized heaps}
\label{sub:cent_heaps_vanish}

We are now prepared to state the main result of Sec.~\ref{sec:heap_freeness}:

The collection of local interaction operators $\{ O^i \}$ with dependency graph $\mathcal{DG}$ are \emph{heap free} in the large $N$ limit. 
That is, the centralized moments of any canonical heap $W^1 \cdots W^{p'}$ vanish:
\begin{align}
\label{eq:heapfreeness}
    \left\ldb (W^1 - \ldb W^1 \rdb) \cdots (W^{p'} - \ldb W^{p'} \rdb) \right\rdb = 0
\end{align}
Since any string of operators built out of the $O^i$ may be commuted into heap form, the vanishing of centralized heaps completely determines all mixed moments recursively in terms of the moments of individual $O^i$.

The proof of heap freeness follows from the representation of the moment on the left of Eq.~\eqref{eq:heapfreeness} as a sum over dependency partitions $\tau$ as in Eq.~\eqref{eq:dp_sum_poly} with $P^i(O^i) = W^i - \ldb W^i \rdb$.
The term corresponding to $\tau$  is proportional to a product over traces of the blocks $B$:
\begin{align}
    \prod_{B \in \tau} \tr \left[ (W^{B_1}-\ldb W^{B_1} \rdb) \cdots (W^{B_{|B|}} - \ldb W^{B_{|B|}}\rdb) \right]
\end{align}
If any block $B\in \tau$ has length 1, the corresponding term vanishes.
Observe that \emph{every} dependency partition $\tau$ arising in the expansion of a heap has a block of length $1$:
Consider the first brick $W^1$. If it lies in a block of length $1$, we are done. If not, then it connects to some brick $W^l$ of the same color with $1 < l \le p$.  
Since the $W$'s form a heap, there must be a dependent brick $W^k$ of another color with $1 < k < l$ -- assume $k$ is the least such brick in the heap. 
Now, repeat this argument starting with the brick $W^k$. 
Either the block containing $W^k$ is of length 1 or it connects to another brick at a position strictly between $k$ and $l$ (because the blocks of dependent colors cannot cross in the dependency partition $\tau$). 
Since the region where the blocks may lie gets strictly narrower at each step, repeating this search must end eventually with a block of length $1$.
QED.

We note that heap freeness subsumes both free independence and classical independence as special cases. If $\{O^i\}$ are freely independent, then $\mathcal{DG}$ is fully connected, and the canonical heaps are alternating, in the sense of  Eq.~\eqref{eq:cent_mom_zero}.
Similarly, if $\{ O^i \}$ are classically independent, then $\mathcal{DG}$ is fully disconnected and the canonical heaps have exactly one brick of any given color $i$. Heap freeness is equivalent to the notion of  $\Lambda$-freeness\cite{Mlotkowski2004}, \emph{a.k.a.} $\epsilon$-freeness\cite{Speicher2016}, developed as an algebraic generalization of free probability. 
From this point of view, our results show that generic Hamiltonians on extended interaction graphs provide a physically motivated realization of a $\Lambda$-free algebra in the large $N$ limit.

\section{Free cumulant expansion}
\label{sec:cumulantexp}

Consider the evaluation of $\ldb \mathrm{A} \mathrm{C} \mathrm{B} \mathrm{C} \mathrm{B} \mathrm{A}^2 \mathrm{C} \mathrm{B} \rdb$ from Fig.~\ref{fig:heap}. 
We can obtain it by summing over all the planar diagrams as in Eq.~\eqref{eq:mono_stacked_planar} or recursively using heap freeness with Eq.~\eqref{eq:heapfreeness}. 
However, both these methods involve tedious algebra which can be avoided by using an alternative formulation of the moments in terms of free cumulants.

Let us briefly review the combinatorial definition of free cumulants. 
The $p$'th free cumulant $\kappa_p$ of a random variable $O$ is defined implicitly by an expansion over non-crossing partitions,
\begin{align}
    \ldb O^p \rdb = \sum_{\tau \in NC(p)}\prod_{B \in \tau} \kappa_{|B|}(\underbrace{O,\cdots,O}_{|B|\textrm{ times}})
\end{align}
In terms of connection diagrams representing the partitions, we can write 
\begin{align*}
    \ldb O \rdb &= 
    \tikz[srp]{\node[spart] (O1) {\text{$O$}}}
    \begin{tikzpicture}[srp,overlay]
        \draw [thick,red] ($(O1.north)+(0,3pt)$) -- +(0,4pt);
    \end{tikzpicture} = \kappa_1(O)
    \vbox to 15pt{} \\
    \ldb O^2 \rdb &=
    \tikz[srp]{\node[spart] (O1) {\text{$O$}}}
    \tikz[srp]{\node[spart] (O2) {\text{$O$}}}
    \begin{tikzpicture}[srp,overlay]
        \draw [thick,red] ($(O1.north)+(0,3pt)$) -- +(0,4pt)
        -| ($(O2.north)+(0,3pt)$);
    \end{tikzpicture}
    +
    \tikz[srp]{\node[spart] (O1) {\text{$O$}}}
    \tikz[srp]{\node[spart] (O2) {\text{$O$}}}
    \begin{tikzpicture}[srp,overlay]
        \draw [thick,red] ($(O1.north)+(0,3pt)$) -- +(0,4pt);
        \draw [thick,red] ($(O2.north)+(0,3pt)$) -- +(0,4pt);
    \end{tikzpicture}
    = \kappa_2(O,O) + \kappa_1(O)\kappa_1(O) 
    \vbox to 15pt{} 
\end{align*}
and so on.
Standard cumulants are defined analogously except that the sum in the moment-cumulant expansion runs over all partitions rather than only the non-crossing ones.
The $p$'th free cumulant $\kappa_p$ is actually a multilinear functional on the algebra generated by $O$. 
For more details, see \cite{NicaSpeicher2006}.

One of the central results of free probability theory is that collections of freely independent operators $\{ O^i \}$ satisfy a monochromatic moment-free cumulant expansion,
\begin{equation}
\label{eq:free_cum_exp}
\ldb O^{i_1} O^{i_2} \cdots O^{i_p} \rdb = \sum_{\tau \in \mathrm{MNC}} \prod_{B \in \tau} \kappa_{|B|}(O^{B_1}, \cdots, O^{B_{|B|}})
\end{equation}
where $\mathrm{MNC}$ denotes the set of monochromatic non-crossing partitions of the $p$ operators. 
This follows from the general recursive definition of the free cumulant $\kappa_k$,
\begin{equation}
\kappa_k (O^{i_1},...,O^{i_k}) = \ldb O^{i_1}...O^{i_k} \rdb - \sum_{\tau \in \mathrm{NC}(p)} \prod_{B \in \tau} \kappa_{|B|}(O^{B_1}, \cdots, O^{B_{|B|}})
\end{equation}
as a sum over \emph{all} non-crossing partitions and the theorem that mixed (ie. multicolor) free cumulants vanish for freely independent operators.
This theorem plays a role precisely analogous to the vanishing of mixed cumulants for classically independent variables and leads to, for example, algebraic proofs of a (free) central limit theorem. 

For a given collection of operators $\{O^i\}$ which are heap free with respect to $\mathcal{DG}$, arbitrary mixed moments $\ldb O^{i_1} \cdots O^{i_p} \rdb$ may be expanded in terms of monochromatic free cumulants by first bringing them to canonical heap form $\ldb W^1 \cdots W^{p'} \rdb$ and then summing over monochromatic \emph{dependency} partitions:
\begin{align}
\label{eq:dp_cum_exp}
\ldb W^{1}W^{2}\cdots W^{p'} \rdb = \sum_{\tau \in \DP} \prod_{B \in \tau} \kappa_{|B|}(W^{B_1}, \cdots, W^{B_{|B|}})
\end{align}
Although the sum over dependency partitions appears naively similar to that in Eq.~\eqref{eq:dp_sum}, the free cumulants  on the right are not the same as the trace moments in Eq.~\eqref{eq:dp_sum}. 
Indeed, the cumulant expansion has several calculational advantages: 
there are no complicated coefficients $C_\tau$ depending on $\mathcal{DG}$. Additionally, for operators with semi-circle law distributions, all of the free cumulants of order $k>2$ vanish, further simplifying the expansion.

For example, let $A,B,C$ have an open chain as their dependency graph (Eq.~\eqref{eq:abc_depgraph}) and satisfy a semi-circle law so that $\kappa_1=0, \kappa_2 = 1, \kappa_{k>2}=0$. 
Equation~\eqref{eq:dp_cum_exp} leads to easy evaluation of correlators such as
\tikzstyle{srp}+=[baseline,remember picture]
\tikzstyle{spart}=[anchor=base,use as bounding box, inner sep=0pt, outer sep=0pt]
\begin{align}
    \ldb ACB^2ACB^2 \rdb &= 
    \tikz[srp]{\node[spart] (A1) {\text{$A$}}}
    \tikz[srp]{\node[spart] (C1) {\text{$C$}}}
    \tikz[srp]{\node[spart] (B1) {\text{$B$}}}^2
    \tikz[srp]{\node[spart] (A2) {\text{$A$}}}
    \tikz[srp]{\node[spart] (C2) {\text{$C$}}}
    \tikz[srp]{\node[spart] (B2) {\text{$B$}}}^2
    \begin{tikzpicture}[srp,overlay]
        \draw [thick,color=red] ($(A1.north)+(0,3pt)$) -- +(0,8pt) -| ($(A2.north)+(0,3pt)$);
        \draw [thick,color=green] ($(C1.north)+(0,3pt)$) -- +(0,11pt) -| ($(C2.north)+(0,3pt)$);
        \draw [thick,color=blue] ($(B1.north)+(0,3pt)$) -- +(0,4pt);
        \draw [thick,color=blue] ($(B2.north)+(0,3pt)$) -- +(0,4pt);
    \end{tikzpicture}
    \vbox to 22pt{} \nonumber\\
    &=\kappa_2(A, A) \kappa_2(C, C) (\kappa_1(B^2))^2=1
\end{align}
As another example, the heap in Fig.~\ref{fig:heap}, $\ldb \mathrm{A} \mathrm{C} \mathrm{B} \mathrm{C} \mathrm{B} \mathrm{A}^2 \mathrm{C} \mathrm{B} \rdb$, vanishes because it admits no monochromatic dependency partitions without length 1 blocks.

The proof of the moment-cumulant formula for heap free operators, Eq. \eqref{eq:dp_cum_exp}, follows from the observation that the RHS vanishes if the heap is centralized. 
Indeed, since $W^1 \cdots W^{p'}$ is a canonical heap, every dependency partition $\tau$ contains some block of length $1$ (by the same argument sketched in Sec.~\ref{sub:cent_heaps_vanish}).
But $\kappa_1(W^j) = \ldb W^j \rdb = 0$ if $W^j$ is centralized; thus, all of the terms in the RHS vanish. 
Since the vanishing of centralized heaps and linearity completely determine all moments recursively, the LHS and the RHS must coincide.

Although Eq.~\eqref{eq:dp_cum_exp} expands mixed moments in terms of (monochromatic) free cumulants, it is important to note that mixed free cumulants of the $\{W^i\}$ need not vanish (unless $\mathcal{DG}$ is fully connected, i.e., all the operators are freely independent with respect to each other).
It is possible to define a $\mathcal{DG}$-dependent cumulant functional for which the mixed cumulants vanish \cite{Speicher2017}, but it is technically more challenging than the direct proof of Eq.~\eqref{eq:dp_cum_exp} sketched here.

\section{Quantum Satisfiability at large-N}
\label{sec:qsat}

As a non-trivial application of the theory developed here, we turn to studying quantum satisfiability (QSAT) at large $N$. 

The QSAT problem generalizes classical constraint satisfaction problems to a quantum setting\cite{Bravyi2011}:
Does the Hamiltonian
\begin{align}
\label{eq:hamqsat}
H = \sum_{m=1}^{M} \Pi_m
\end{align}
have a zero energy (\emph{satisfying}) ground state? 
Here, each projector $\Pi_i$ represents a constraint which must be satisfied by the subset of $n$ qudits on which it acts, according to a given interaction graph $\mathcal{G}$.
Deciding whether $H$ is satisfiable is QMA$_1$-complete\cite{Gosset2016} and thus expected to be algorithmically intractable even with the aid of a quantum computer. 

On the other hand, much progress can be made on \emph{generic} QSAT. 
After fixing the discrete data describing a QSAT instance -- the interaction graph $\mathcal{G}$, dimension of the qudits $N_q$ and relative ranks of the projectors $p_i \equiv R(\Pi_i) = \tr \Pi_i$ -- the \emph{geometrization theorem} \cite{Laumann2010} asserts that almost all choices of the $\Pi_i$ produce a minimal SAT dimension, $R(\ker H)$. 
That is, generic QSAT instances are as frustrated as possible and we can study average behavior in the Haar randomized projector model in order to uncover the worst case. 


The large $N_q = N$ limit is non-trivial on any given graph $\mathcal{G}$ so long as the relative ranks $p_i$ are held fixed as the limit is taken. It has been conjectured\cite{Sattath2016} that in this limit, 
\begin{align}
    R( \ker H )= \mathcal{Z}(\mathcal{DG}, -p)
\end{align}
where $\mathcal{Z}$ is the partition function for a classical  hard-core lattice gas of particles living on the dependency graph $\mathcal{DG}$ at fugacity $-p$. 
For the computer science oriented reader, $\mathcal{Z}$ is also called the Shearer\cite{Shearer1985} or independent set polynomial.

In the large-$N$ limit, the projectors $\Pi_i$ become heap free with respect to $\mathcal{DG}$ and, in principle, all of the moments of $H$ are determined.
To solve QSAT, one needs to compute all of those moments $\ldb H^k \rdb$ to reconstruct the spectral weight of $H$ at $E=0$ corresponding to the satisfying space.
This is possible if one has an analytic approach to managing the calculation, for example through convolution theorems and generating functions.
Unfortunately, we have not been able to discover a general convolution theorem for heap free variables.

\begin{figure}[tbp]
\begin{center}
\includegraphics{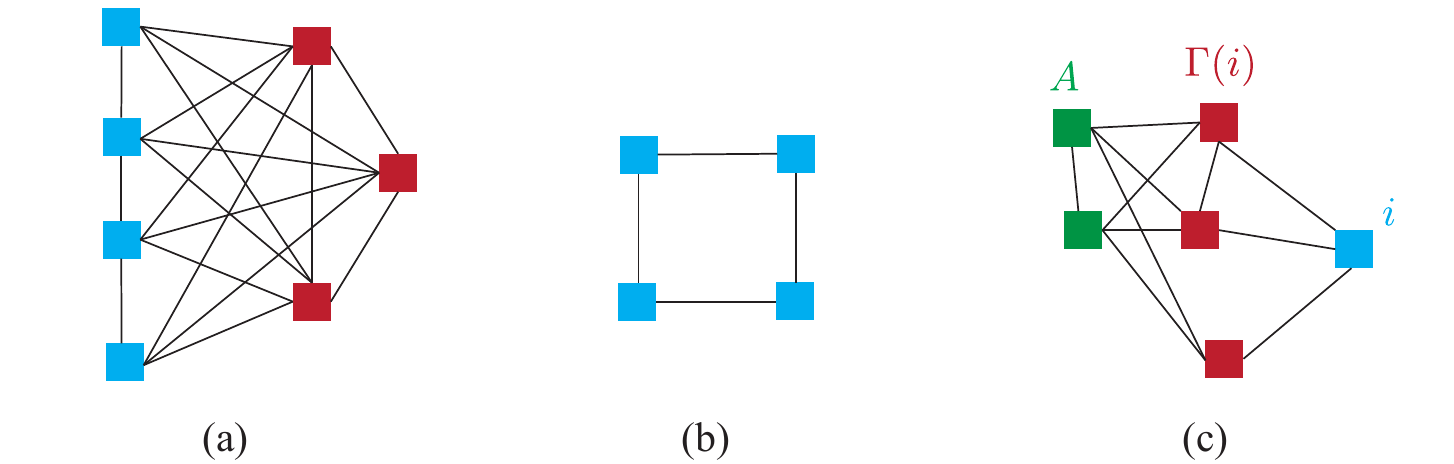}
\caption{
(a) A dependency graph which is bipartite fully connected, i.e., every blue box is connected to every red box. If the zero energy space of the blue terms and the red terms are known, then the free sum rule provides the zero energy space of the full system.
(b) The zero energy space of the 4-cycle can be solved. 
(c) Decomposition of a dependency graph into a site $i$, its neighbors $\Gamma(i)$ and the rest of the system $A$ with $\Pi_{\Gamma(i)}$ and $\Pi_A$ satisfying the free sums property. 
}
\label{fig:graphsQSAT}
\end{center}
\end{figure}

Nonetheless, progress can be made for a large class of dependency graphs $\mathcal{DG}$ (such as those in Fig.~\ref{fig:graphsQSAT}) using the following properties
\begin{itemize}
    
    \item (classical combination) If the subset of operators $\{\Pi^i\}$ are disconnected in $\mathcal{DG}$, then $I=\sum_i \Pi^i$ has zero energy space with relative dimension $K = \prod_i (1-p_i)$.

    \item (free sums) If there exist two subsets of operators $\{\Pi^i\}$ and $\{\Pi^j\}$ such that every operator in the first set is connected to every operator in the second set (in $\mathcal{DG}$), then $I = \sum_i \Pi^i$ and $J = \sum_j \Pi^j$ are freely independent variables.
    
    \item (free combination) If $I$ and $J$ have zero spaces with relative dimension $K_I = 1-p_I$ and $K_J = 1-p_J$ respectively, then $I+J$ has zero energy space with relative dimension $K_{I+J} = \frac{1}{2}(1 - p_I - p_J + |1-p_I-p_J|)$.
\end{itemize}

The classical combination property is trivially true since the resulting operators are tensor independent and the combined spectrum follows from a convolution corresponding to classical random variables.

The free sums property follows from the formalism developed so far. To prove that $I$ and $J$ are free, we should show Eq.~\ref{eq:cent_mom_zero} holds where the operators are alternatively functions of $I$ and $J$. By linearity, it is enough to show that centralized moments which involve combinations of individual operators from $I$ and $J$ alternatively are zero. Since every operator in the first set is connected to every operator in the second set in $\mathcal{DG}$, the connections of the first set are non-crossing with respect to those of the second set. Thus, from the arguments given in Sec.~\ref{sec:heap_freeness}, there is at least one block of length one which implies centralized moments vanish.

To prove the free combination property, we first note that quantum satisfiability reduces to determining whether the spectrum, $\rho(z) = \ldb \delta(z - H) \rdb$, has weight at $E=0$.
Equivalently, we are interested in whether the resolvent
\begin{align}
    G(z) &= \left\ldb \frac{1}{z - H} \right\rdb = \frac{K}{z} + \cdots
\end{align}
has a pole at $z=0$ with residue $K > 0$. 




Now consider $H=I + J$ where $I$ and $J$ satisfy the second property. Since the spectrum of each Hamiltonian is semi-positive and the rank is invariant under deformations of the positive part of the spectrum which preserve the total weight, we can choose $I$ and $J$ to be a projectors with relative ranks $p_I$ and $p_J$ respectively. 
The associated deformed resolvents are
\begin{align}
G_{I} (z) = \frac{1 - p_I}{z} + \frac{p_I}{z - 1}\\
G_{J} (z) = \frac{1 - p_J}{z} + \frac{p_J}{z - 1}
\end{align}

Since $I$ and $J$ are freely independent, the vanishing of the mixed free cumulants implies that the free cumulants of $I$ and $J$ are additive.
This can be summarized by the additivity of the free cumulant generating function,
\begin{align}
    R_H(w) &\equiv \sum_{k=1}^\infty \kappa_k(H,\cdots,H) w^{k-1}
\end{align}
We recall that $R(w)$ is related to $G(z)$ by the $R$-transform\cite{NicaSpeicher2006},
\begin{align}
\label{eq:GR}
    G(R(w)+\frac{1}{w}) = w
\end{align}
This gives us a complex analytic tool with which to extract the pole in $G_H$ at $z=0$.

Inverting $G_I$ and using that $G^{-1} (g) \approx 1/g $  for $g \in \mathbb{R}$  and $g \rightarrow 0$ (where $g \equiv G(z)$), we get 
\begin{equation}
R_{I} (z) = \frac{z - 1+\sqrt{1  + z (z + 4p_A - 2)}}{2z}
\end{equation}
and a similar equation for $R_{J}$.
Since the free cumulants are additive, we get $R_{I + J} (z) = R_I + R_J$ and hence $G_{I + J} (z) $ from solving Eq.~\ref{eq:GR}. 
After some algebra, the residue of the pole at $z=0$ is given by
\begin{align}
K_{I+J} = \dfrac{1 - p_I - p_J + \lvert 1 - p_I - p_J \rvert}{2} 
\end{align}

If $\mathcal{DG}$ can be recursively decomposed using the above three properties, then we can compute the dimension of the zero energy space of the resulting Hamiltonian. In particular, consider a dependency graph $\mathcal{DG}$ which can be split into a fixed vertex $i$, its neighbors $\Gamma(i)$ and the rest of the system $A$ such that $\Pi_{\Gamma(i)}$ and $\Pi_A$ satisfy the second property (Fig.~\ref{fig:graphsQSAT}(c)). Assume the relative kernels of these sub-systems are known and are equal to the Shearer polynomials $1 - p, \mathcal{Z}_{\Gamma(i)}, \mathcal{Z}_A$ on them respectively and $\mathcal{Z}_{\Gamma(i) + A}$ for the combined system of $\Gamma(i)$ and $A$. Since $\Pi_i$ and $\Pi_A$ are disconnected, we have $\mathrm{ker}(\Pi_{i} + \Pi_{A}) = (1 - p) \mathcal{Z}_A$. From the free sums property, we have that $\Pi_{i} + \Pi_A$ and $\Pi_{\Gamma(i)}$ are free. From the free combination property, we have that $\mathrm{ker}(\Pi_{i} + \Pi_{A}) = \mathcal{Z}_{\Gamma(i) + A} = \mathcal{Z}_A + \mathcal{Z}_{\Gamma(i)} - 1$. Thus, the kernel of the full system (when it is positive) is given by 

\begin{align*}
\mathrm{ker}(\Pi_i + \Pi_{\Gamma(i)} + \Pi_A) &= 1 - p_I - p_J \\
&= 1 - (1 - \mathcal{Z}_{i + A}) - (1 - \mathcal{Z}_{\Gamma(i)}) \\
&= (1 - p) \mathcal{Z}_A + \mathcal{Z}_{\Gamma(i)} - 1 \\
&= \mathcal{Z}_{\Gamma(i) + A} - p \mathcal{Z}_A \\
&= \mathcal{Z}_{i + \Gamma(i) + A}
\end{align*}
Here we have used the recursion relation for the Shearer (independent set) polynomial in the last line. Thus, the kernel for the combined system reduces to the Shearer polynomial of the system at large-$N$.

In particular, the Shearer polynomial/hard-core lattice gas partition function is the exact answer for the dependency graph of a 4-cycle(Fig.~\ref{fig:graphsQSAT}(b)). It has been shown that the classical Shearer theorem is not tight in this setting\cite{Kolipaka2011} so this constitutes a provable separation between quantum and classical constraint satisfaction problems.

We note a recent work\cite{TightShearer2018} shows that the critical relative rank $p_c$ beyond which quantum satisfiability on a fixed dependency graph $\mathcal{DG}$ can be made unsatisfiable is tightly lower bounded by the least zero of the Shearer polynomial.
This does not show that the dimension of the kernel is tightly lower bounded by the Shearer polynomial in the satisfiable regime, as we have shown here.

\section{Discussion and open questions}

We have presented three methods for calculating disorder averaged trace correlators of random Hamiltonian systems with spatial locality at large local Hilbert space dimension $N$. 
The stacked diagram expansion organizes the contributions to such moments by the Euler characteristic $\xi$ of a stack of $n$ 2D layers. 
The leading terms in $\ldb \cdots \rdb=\overline{\tr \cdots}$ correspond to stacked planar diagrams; the higher genus corrections vanish as $N\to \infty$.
The Euler characteristic expansion further shows that the trace moments themselves factorize,
\begin{align}
    \overline{\tr W^1 \tr W^2} = \overline{\tr W^1}\,\,\overline{\tr W^2} + O\left(\frac{1}{N}\right)
\end{align}
for any operators $W^1, W^2$ constructed from the $O^i$. 
This follows because each trace provides additional circular boundaries which reduce the Euler characteristic of the connected diagrams. 
This is a form of concentration of measure for large $N$.

In the strictly planar limit, we have shown that the operators $O^i$ become heap free with respect to the dependency graph $\mathcal{DG}$.
This combinatorial result lead us to several compact methods of organizing the calculation of the average correlators.
The most powerful of these expresses the average moments as a sum over dependency partitions of products of free cumulants.
This also connects the physical theories described by Hamiltonian \eqref{eq:generalham} to the recent generalizations of free probability theory to incorporate mixed collections of commuting and non-commuting operators.

There are many open questions and directions to pursue building on this work. 

\begin{enumerate}
    \item In random matrix theory, the $1/N$ corrections are universal in the sense that they encode level repulsion and quantum dynamics on very long time scales. It would be very interesting to show level repulsion explicitly in the fluctuation corrections to the full many-body Hamiltonian spectrum of an extended system. This would operate on an energy scale $1/N^n$ corresponding to the many-body level spacing.
    \item How do the results change for random local many-body Hamiltonians with different symmetry groups?
    \item Hamiltonians of the form Eq.~\eqref{eq:generalham} on finite dimensional lattices should exhibit both energy diffusion and scrambling of quantum operators. Demonstrating these explicitly and studying their interplay would be of great interest.
    \item As mentioned previously, the all-to-all version of Eq.~\eqref{eq:generalham} maps onto a fermionic model closely related to the Sachdev-Ye-Kitaev models. Whether the techniques in this paper can be used to provide complementary information regarding these models is an intriguing future direction.
     \item Is it possible to construct a generalized convolution theorem to obtain the spectrum of a sum of terms involving a mixture of classically independent and freely independent variables with a specified dependency graph? It is easy to see that it cannot be a binary operation but a ternary operation is conceivable.
    \item Does the independence polynomial provide the relative dimension of the quantum satisfying space for QSAT at large $N$ for all graphs $\mathcal{G}$?
\end{enumerate}

\section*{Acknowledgements}
We thank C.L. Baldwin, G. C{\'e}bron, J. Chalker, A. Chandran, R. Moessner, O. Sattath, R. Speicher and T. Tao for stimulating conversations and feedback on this work at various stages. 
CRL acknowledges support from the Sloan Foundation through a Sloan Research Fellowship and the NSF through CAREER grant No. PHY-1752727. 
Any opinion, findings, and conclusions or recommendations expressed in this material are those of the authors and do not necessarily reflect the views of the NSF.

\bibliographystyle{unsrt}
\bibliography{refs}

\end{document}